# Two kinds of robustness are not the same: disentangling fault tolerance and low-SNR robustness in multi-domain event detection on real data

Isao Kurosawa

*IVXA, Japan*

Corresponding author: Isao Kurosawa (IVXA, Japan).

## Highlights

- Sensor-loss and low-SNR robustness are distinct properties with distinct causes.
- A multi-domain real benchmark spans seismic, borehole DAS, and industrial vibration.
- Low-SNR robustness arises from sensor-dropout training, not architectural redundancy.
- Same-architecture ablations isolate the mechanism (recipe-off vs redundancy-off).
- Simple detectors with sensor-dropout deliver much field robustness cheaply.

## Abstract

Reliable event detection underpins induced-seismicity monitoring for Carbon dioxide Capture and Storage (CCS) and geothermal operations, distributed acoustic sensing (DAS), and industrial condition monitoring. In each setting a detector must stay reliable both when sensors fail and when the signal is buried in noise. These two failure modes are routinely conflated, and architectural complexity is often credited with robustness it may not deserve. We assemble a unified binary event-detection benchmark from three physically distinct real sources -- Hi-net seismic waveforms, Utah FORGE 2024 borehole DAS, and MAFAULDA industrial vibration -- each mapped to a common 8-channel, 256-sample representation, and evaluate a fault-tolerant detector (CEPHALON) trained with per-sample sensor-dropout against standard detectors (a 1D convolutional network, a temporal convolutional network, and a compact Transformer) trained with an identical recipe. On clean data every model is near-perfect (AUC ~ 0.99). Under progressive sensor loss, simple models with sensor-dropout are already robust and CEPHALON holds no advantage. Under additive noise, however, CEPHALON degrades far more gracefully: at -2.5 dB its overall AUC is 0.939 versus 0.532-0.572 for the convolutional baselines. Same-architecture ablations isolate the cause: disabling internal redundancy at inference reduces the low-SNR advantage only modestly, whereas removing sensor-dropout training collapses it (0.899 to 0.603 at -5 dB). The training recipe is therefore the dominant cause and parallel redundancy only secondary. We release a complete, numbered, reproducible pipeline so that every figure can be regenerated.

**Keywords:** distributed acoustic sensing; induced seismicity; event detection; robustness; data augmentation; deep learning

## 1. Introduction

Continuous monitoring of induced seismicity has become a routine operational requirement for Carbon dioxide Capture and Storage (CCS) and for geothermal energy, where injection and production perturb the local stress field and can reactivate faults (Ellsworth, 2013; Grigoli et al., 2017). Reliable, automated detection of small events is the first link in every downstream task, from catalogue construction to hazard management, and distributed acoustic sensing (DAS) has made dense, borehole-deployed observation practical by turning a single optical fibre into thousands of co-located strain-rate sensors. Learned detectors have become standard for such tasks, replacing hand-tuned characteristic functions with data-driven feature extractors. Yet their robustness is rarely dissected along independent axes, even though field deployments expose detectors to exactly the conditions under which laboratory accuracy is least informative.

A field detector is always embedded in a sensor network that is noisy and only partially reliable. Channels drop out, optical coupling varies along the fibre, instruments are decommissioned, and the events of interest are frequently small relative to background noise. A useful detector must therefore degrade gracefully along two independent axes: when sensors are lost, and when the signal-to-noise ratio (SNR) falls. These two axes are easy to conflate, but they are not the same. A model that survives the loss of several input channels is not necessarily a model that survives heavy additive noise, and a model that tolerates noise need not tolerate missing sensors. The two stresses act on different parts of the inference problem — one removes information channels, the other corrupts the information within them — and there is no a priori reason for a single architectural or training choice to address both equally.

Robustness claims in the monitoring literature nevertheless bundle the two together, and improvements are commonly attributed to architectural ingredients — redundant pathways, mixtures of experts, attention — without a causal test of whether the architecture, rather than the training procedure, is responsible. The distinction matters in practice. If robustness is conferred by the training recipe, it is portable and cheap: any detector, including a lightweight one suited to edge deployment on an acquisition node, can inherit it. If robustness is conferred by a heavy, redundant architecture, it carries a permanent computational cost. On real, multi-domain data this question has rarely been isolated, in part because most studies evaluate a single modality and a single architecture, leaving recipe and architecture confounded.

We address the question directly. We assemble a unified event-detection benchmark from three physically distinct real datasets and stress-test detectors along both robustness axes, comparing a deliberately redundant, bio-inspired architecture (CEPHALON) against standard baselines trained with the same recipe. Holding the training recipe fixed across architectures, and holding the architecture fixed across training recipes, lets us separate the two contributions causally rather than correlationally. Our contributions are: (1) a unified, leak-free, multi-domain real-data benchmark spanning seismic, borehole DAS, and industrial vibration at a common 8-channel, 256-sample representation; (2) a two-axis robustness evaluation — progressive sensor loss and progressive low SNR — against strong, identically trained baselines; (3) a causal disentangling of the two robustness types, showing that the low-SNR advantage is driven by sensor-dropout training, not by parallel architectural redundancy; and (4) a practical takeaway for field monitoring, delivered as a fully reproducible, numbered pipeline. We also report negative and null results honestly, including that simple detectors match the redundant architecture under sensor loss and that zero-shot cross-domain transfer fails while few-shot adaptation succeeds.

The remainder of this paper is organized as follows. Section 2 reviews learned event detection, machine learning for DAS, and robustness and data augmentation in deep learning. Section 3 describes the three datasets and the unified representation. Section 4 details the CEPHALON architecture, the sensor-dropout training recipe, the identically trained baselines, and the evaluation protocols. Section 5 reports the clean, sensor-loss, low-SNR, mechanism, and transfer results. Section 6 discusses why the recipe transfers across corruption types and what this implies for field monitoring, and Section 7 concludes.

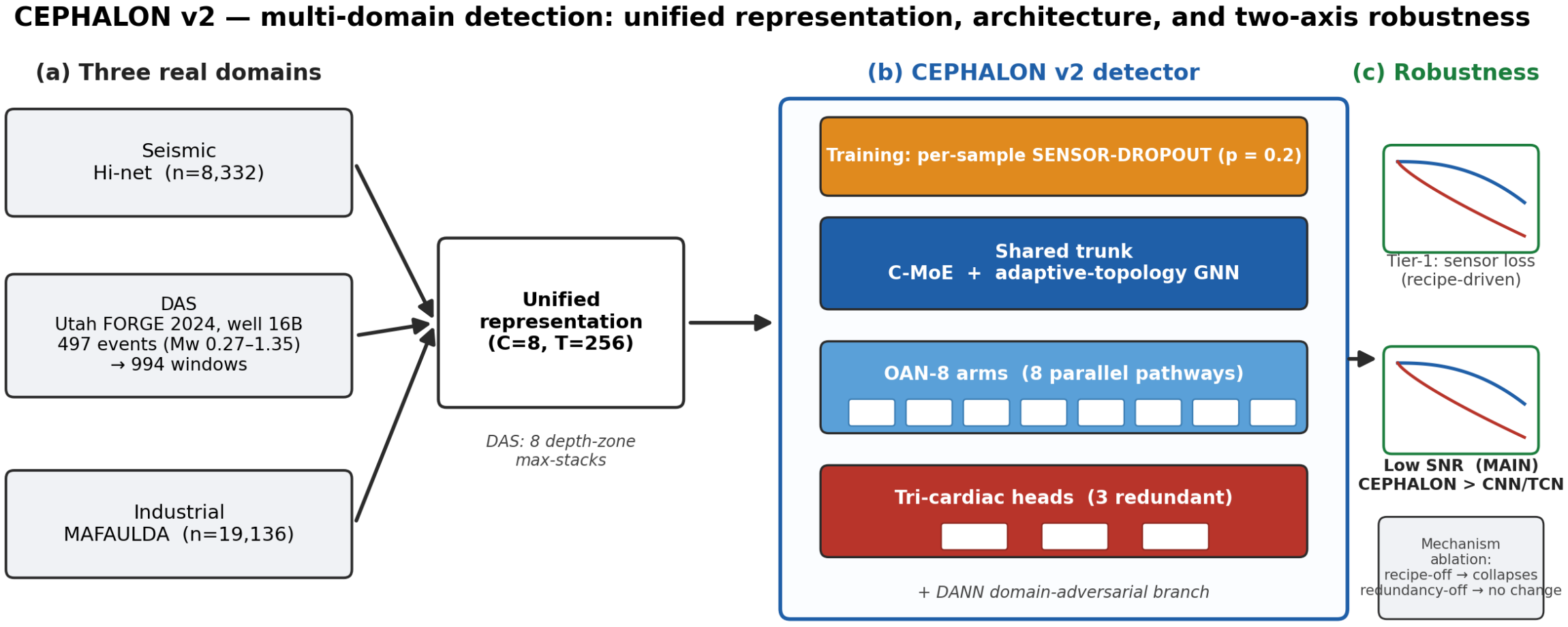

**Fig. 1.** Overview. Three real domains are reduced to a common (C = 8, T = 256) representation; CEPHALON (shared trunk, OAN-8 arms, tri-cardiac heads, DANN branch) is trained with per-sample sensor-dropout; a detector is stress-tested along two robustness axes, and same-architecture ablations isolate a mechanism.

## 2. Related work

### *2.1 Learned event detection for seismicity monitoring*

Deep learning has reshaped automatic event detection and phase picking in seismology (Mousavi and Beroza, 2022). Convolutional detectors such as ConvNetQuake (Perol et al., 2018) and generalized phase detection (Ross et al., 2018) demonstrated that learned feature extractors can match or exceed classical short-term-average / long-term-average triggers, while PhaseNet (Zhu and Beroza, 2019) established U-shaped fully convolutional picking as a community standard. Attention-based models such as the Earthquake Transformer (Mousavi et al., 2020) jointly perform detection and picking and generalize across networks. These advances target detection accuracy on relatively complete, well-coupled networks; comparatively little attention has been paid to how such detectors behave when the input network is degraded by missing channels or elevated noise, which is precisely the operating regime of dense, automated monitoring around CCS and geothermal sites.

### *2.2 Distributed acoustic sensing and machine learning for borehole monitoring*

DAS interrogates an optical fibre to recover distributed strain or strain-rate along its length, providing thousands of densely spaced channels suitable for borehole and surface deployment (Lindsey and Martin, 2021). The modality is attractive for induced-seismicity monitoring because a single fibre can instrument an entire well at reservoir depth, but it raises distinctive challenges: spatially varying coupling, strong instrument and optical noise, a directional (axial-strain) sensitivity that differs from particle-velocity seismometers, and very large data volumes. Machine learning has been applied to DAS denoising and coherence enhancement (van den Ende et al., 2021) and to phase picking and event detection (Zhu et al., 2023), yet most studies treat all channels as available and clean. The combination of dense channels and variable coupling makes DAS an ideal modality for studying the two robustness axes we isolate here, because both channel loss and low SNR occur naturally along a fibre.

### *2.3 Robustness, data augmentation, and fault tolerance in deep learning*

Robustness to input corruption is a long-standing concern in machine learning. Benchmarks of common corruptions and perturbations (Hendrycks and Dietterich, 2019) show that accuracy on clean data is a poor predictor of accuracy under distribution shift, and that targeted augmentation often improves corruption robustness more than architectural changes. Structured input perturbations — dropout (Srivastava et al., 2014), spatial masking such as Cutout (DeVries and Taylor, 2017) — act as regularizers that discourage over-reliance on any single feature or region. Our sensor-dropout is a domain-specific instance of this family applied at the channel (sensor) level. Fault tolerance, by contrast, is usually pursued architecturally, through redundant or ensembled pathways. A central question, rarely tested on real geophysical data, is whether observed robustness is attributable to the recipe (the augmentation) or to the architecture (the redundancy). We answer it with same-architecture ablations.

### *2.4 Multi-domain learning and condition monitoring*

Casting heterogeneous sensing problems into a shared representation invites multi-domain and domain-adaptation techniques. Domain-adversarial training (Ganin et al., 2016) encourages domain-invariant features through a gradient-reversal branch, and sparsely gated mixtures of experts (Shazeer et al., 2017) allocate capacity conditionally. Rotating-machinery fault diagnosis, exemplified by the MAFAULDA database (Ribeiro, 2018; Marins et al., 2018), provides a vibration-monitoring domain whose statistics differ sharply from seismic and DAS data, making it a useful third pole for a multi-domain benchmark. We do not claim cross-domain transfer as a strength; on the contrary, we report that zero-shot transfer across these physically distinct domains fails, and that the value of the shared representation is realized only through few-shot adaptation.

## 3. Data and unified representation

### *3.1 Unified binary detection task and representation*

We construct a single binary “event present / absent” task shared across three domains. Every window from every domain is mapped to a common tensor (Eq. 1)

$$x \in \mathbb{R}^{C\times T}, \quad C = 8, \quad T = 256, \tag{1}$$

so that a single detector and a single set of baselines can be trained and stressed identically across modalities. The choice of eight channels reflects the smallest sensor count that still admits meaningful sensor-loss and component-loss experiments, and 256 samples provides a window long enough to contain a complete short event after decimation. Class labels are binary and a separate domain label $d \in \{0, 1, 3\}$ (seismic, DAS, industrial) is retained for the domain-adversarial branch and for per-domain reporting.

### *3.2 Seismic domain (Hi-net)*

For the seismic domain ($d = 0$) we use waveform windows from the NIED High-Sensitivity

Seismograph Network, Hi-net (Okada et al., 2004), yielding 8,332 class-balanced examples. Windows are reduced to the common eight-channel representation and normalized so that the amplitude cue is preserved while the overall scale is bounded across domains.

### *3.3 Borehole DAS domain (Utah FORGE 2024)*

For the DAS domain (d = 1) we use Utah FORGE 2024 borehole DAS acquired by Neubrex in monitoring well 16B during the April 2024 stimulations (Ajo-Franklin et al., 2024; Utah FORGE GDR, 2024). From the triggered-file catalogue we take 497 events spanning Mw 0.27–1.35 (median 0.65). Each triggered file contains one catalogued event, which we locate within the file by maximum band energy rather than by the unreliable catalogue offset; a quiet interior window is taken as the paired noise example, giving a balanced positive/negative split. The fibre band that illuminates the events, channels 1116–1493, is identified automatically by an event-versus-noise band-RMS diagnosis; this deep band is consistent with reservoir-depth illumination in the 16B well. Surviving channels are reduced to eight contiguous depth-zone stacks (maximum over channels within each zone), band-passed 5–120 Hz, and decimated from 4000 to 500 Hz, producing 994 windows. The split is leak-free by file (train 694 / validation 150 / test 150), so that windows from a single triggered file never appear in more than one split.

### *3.4 Industrial domain (MAFAULDA)*

For the industrial domain (d = 3) we use vibration windows from the MAFAULDA machinery-fault database (Ribeiro, 2018; Marins et al., 2018), yielding 19,136 examples. This domain is deliberately included as a statistically distant third pole: its positive-class prior is high (0.795) and its signal character differs markedly from the seismic and DAS domains, which stresses any claim of domain invariance.

### *3.5 Multi-domain assembly, splits, and normalization*

The three sets are merged into a single benchmark of 28,462 windows with per-domain splits (Table 1). A noise-relative normalization is applied per window, preserving the amplitude cue that distinguishes events from noise while bounding the scale across domains so that no single modality dominates training by magnitude alone. Positive-class priors differ across domains — balanced for DAS, near-balanced for seismic, and high for industrial — so all detection performance is reported with AUC and per-domain threshold calibration, which is insensitive to prior shift. Quality-control diagnostics for the assembled benchmark are shown in Fig. 2.

**Table 1.** Multi-domain dataset composition and leak-free splits. Counts are windows; positive-class prior is the test-split fraction with label 1.

| Domain | Source | Windows | Train / Val / Test | Pos. prior |
|---|---|---|---|---|
| Seismic (d = 0) | Hi-net (NIED) | 8,332 | 5832 / 1250 / 1250 | ≈ 0.39 |
| DAS (d = 1) | Utah FORGE 2024, well 16B | 994 | 694 / 150 / 150 | 0.50 |
| Industrial (d = 3) | MAFAULDA | 19,136 | 13440 / 2848 / 2848 | 0.80 |
| **Total** | — | **28,462** | — | — |

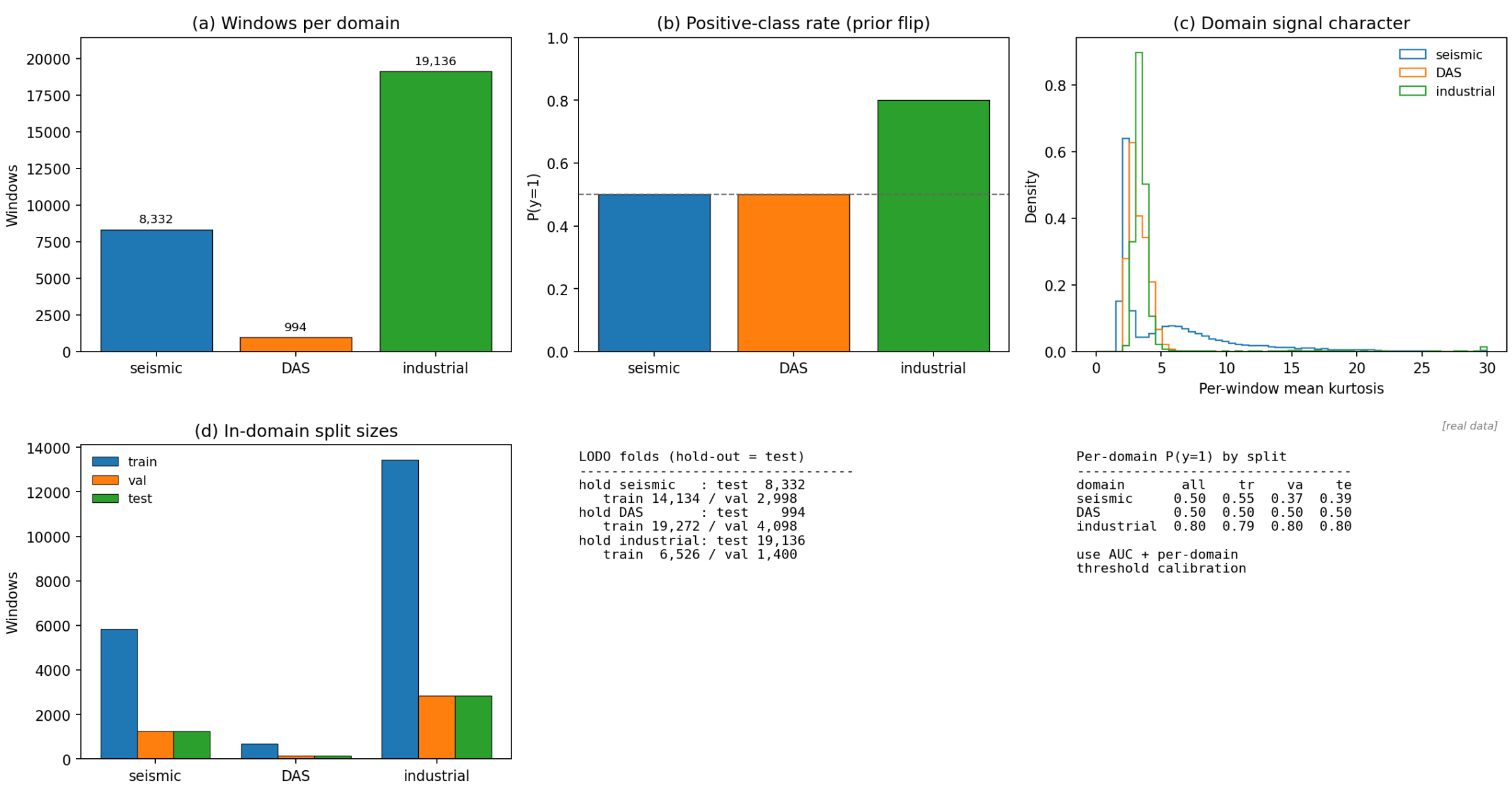

**Fig. 2.** Multi-domain dataset assembly quality control: window counts per domain, positive-class rate (prior flip across domains), per-window signal character, and in-domain split sizes.

# 4. Methods

## *4.1 CEPHALON architecture*

CEPHALON is a multi-domain detector composed of a shared trunk, eight parallel arms, three redundant classification heads, and a domain-adversarial branch. The design is deliberately and visibly redundant so that the contribution of redundancy can be measured directly. The single design choice central to this paper, however, is not the redundancy but the training-time augmentation described in Section 4.2.

**Shared trunk.** The trunk combines a channel mixture-of-experts (C-MoE), in the sparsely gated form of Shazeer et al. (2017), with an adaptive-topology graph module in the spirit of graph convolutional networks (Kipf and Welling, 2017). For an input x the C-MoE computes a gate over E experts from a pooled summary of the input and forms a conditionally weighted combination of expert outputs (Eq. 2),

$$g = \mathrm{softmax}(W_g \cdot \mathrm{pool}(x)), \quad z = \Sigma_e\, g_e\, E_e(x), \tag{2}$$

where $E_e$ is the e-th expert and the gate g concentrates capacity on the experts most useful for each window. An adaptive adjacency A over the eight channels is learned from node features and used to propagate information across channels (Eq. 3),

$$A_{ij} = \sigma(\varphi(h_i, h_j)), \quad H' = \sigma(\hat{A}\, H\, W), \tag{3}$$

with $\hat{A}$ the row-normalized learned adjacency, H the matrix of channel features, W a trainable weight, and σ a nonlinearity. The adaptive topology lets the trunk reweight inter-channel coupling per window rather than assuming a fixed graph.

**OAN-8 arms.** Eight parallel arms (an octopus-inspired arm network, OAN-8) $f_1 \ldots f_8$ process trunk features independently; an arm-alive mask $a \in \{0, 1\}^8$ selects which arms contribute, and surviving arms are pooled (Eq. 4),

$$r = ( \Sigma_k a_k f_k(\cdot) ) / \max(1, \Sigma_k a_k). \quad (4)$$

**Tri-cardiac heads.** Three redundant classification heads (a tri-cardiac module) map the pooled representation to logits; a heart-alive mask $b \in \{0, 1\}^3$ averages over surviving heads (Eq. 5),

$$\ell = ( \Sigma_m b_m \, \text{head}_m(r) ) / \max(1, \Sigma_m b_m) \in \mathbb{R}^2. \quad (5)$$

Because the arms and heads are trained with random alive masks, the network is explicitly prepared to operate with arbitrary arm and heart subsets at inference; the component-loss and redundancy-off conditions in Section 4.5 are therefore in-distribution rather than out-of-distribution perturbations.

**Domain-adversarial branch.** A domain classifier receives trunk features through a gradient-reversal layer (Ganin et al., 2016). During the backward pass the gradient from the domain loss is negated, so the trunk is trained to make features that the detector can use but the domain classifier cannot, encouraging a degree of domain invariance (Eq. 6),

$$\nabla_{\theta\text{trunk}} \leftarrow \nabla_{\theta\text{trunk}} ( L_{det} - \lambda_d L_{dom} ). \quad (6)$$

### *4.2 Sensor-dropout training*

The training-time augmentation that this paper isolates as the causal driver of low-SNR robustness is per-sample sensor-dropout, a structured-corruption augmentation that builds on dropout (Srivastava et al., 2014) but acts at the level of physical sensors rather than hidden units. For each training window an independent Bernoulli mask is drawn over the eight channels (Eq. 7),

$$m_c \sim \text{Bernoulli}(1 - p), \quad p = 0.2, \quad \text{subject to } \Sigma_c m_c \geq 1, \quad (7)$$

and the masked window is presented to the network (Eq. 8),

$$\tilde{x} = m \odot x \quad \text{(broadcast over time)}, \quad (8)$$

where $\odot$ denotes element-wise multiplication and the constraint guarantees that at least one channel is always kept. Each channel is therefore independently zeroed with probability p, teaching the detector not to over-rely on any single sensor and encouraging distributed, redundant feature use. The probability $p = 0.2$ is fixed for all models and all domains. The same augmentation is applied to the baselines in Section 4.4, so that any difference between architectures is not a difference in recipe.

### *4.3 Training objective and optimization*

Detection is trained with a binary cross-entropy loss over the two output logits (Eq. 9),

$$L_{det} = -(1/N) \Sigma_i [ y_i \log \hat{p}_i + (1 - y_i) \log(1 - \hat{p}_i) ], \quad (9)$$

and the total objective adds the gradient-reversed domain-adversarial term (Eq. 10),

$$L = L_{det} + \lambda_d L_{dom}, \quad (10)$$

with L_dom a cross-entropy over the domain label routed through the gradient-reversal layer. Optimization uses Adam (Kingma and Ba, 2015) with a class-balanced sampler. Training selects the best-validation checkpoint (best-epoch checkpointing) so that the reported model is always the best on held-out data rather than the last epoch. Identical sampler, augmentation, optimizer, and checkpointing settings are used for every model in the study.

### *4.4 Identically trained baselines*

To test whether the architecture is necessary we train three standard detectors with the identical

recipe — same balanced sampler, same sensor-dropout ($p = 0.2$), same optimizer, and same best-validation checkpointing: a plain one-dimensional convolutional network (PlainCNN1D), a temporal convolutional network (TCN; Bai et al., 2018), and a compact Transformer encoder (TinyTransformer; Vaswani et al., 2017). All map the common ($C = 8$, $T = 256$) input to two logits. None of the baselines has internal redundant pathways, so the component-loss (Tier-2) condition is structurally unavailable to them; this asymmetry is itself informative and is reported as such.

### *4.5 Robustness evaluation protocols*

Five protocols are used. **(i) Sensor loss:** zero k of the C channels and average test AUC over random k-subsets, sweeping k from 0 to 7. **(ii) Component loss (CEPHALON only):** disable OAN arms and tri-cardiac hearts, sweeping the number of surviving components. **(iii) Low SNR:** add per-window Gaussian noise scaled to a target SNR, with the identical noised test set presented to every model and results averaged over three noise seeds. **(iv) Mechanism ablations:** disable redundancy at inference (1 arm, 1 heart) and, separately, remove sensor-dropout from training (the otherwise identical "vanilla" model). **(v) Cross-domain transfer:** train on two source domains plus K labeled target examples and evaluate on the held-out target (leave-one-domain-out, LODO). Test SNR is defined per window from signal and noise powers (Eq. 11),

$$SNR_{dB} = 10 \log_{10}( P_{signal} / P_{noise} ), \tag{11}$$

and the additive noise is scaled to reach the target SNR before the noised window is presented to every model (Eq. 12),

$$x_{noisy} = x + \alpha n, \quad \alpha \text{ chosen so that } SNR_{dB} = \text{target}. \tag{12}$$

Gaussian (white) noise is used in the main experiments; pink (1/f) noise is used as a colored-noise robustness check in Appendix B.

### *4.6 Evaluation metric and reproducibility*

Detection performance is summarized by the area under the receiver-operating-characteristic curve (AUC; Bradley, 1997), which is threshold-free and insensitive to the per-domain prior shift in Table 1. AUC is reported overall and per domain, and, for the stochastic protocols, as mean ± standard deviation over the noise seeds. All experiments are produced by a numbered pipeline run in a uv environment: ingestion (Hi-net, FORGE 2024 DAS, MAFAULDA), multi-domain assembly, training (in-domain, leave-one-domain-out, few-shot), fault-tolerance and low-SNR evaluation, baseline training, and figure generation. FORGE 2024 triggered data are retrieved from the Geothermal Data Repository data lake using the catalogue keys. All figures use a white background, English labels, and legends placed in the margins.

## 5. Results

### *5.1 Clean in-domain performance*

With all sensors present and no added noise, every model is near-perfect (Table 2). CEPHALON reaches an overall AUC of about 0.999, with per-domain seismic, DAS, and industrial values all close to 1.00, and the baselines fall in the range 0.99–1.00. Differences between models therefore appear only under stress, which motivates the two-axis evaluation that follows.

**Table 2.** Clean in-domain detection performance (AUC), all sensors present, no added noise.

| Model | Overall | Seismic | DAS | Industrial |
|---|---|---|---|---|
| CEPHALON v2 (csd) | ≈ 0.999 | ≈ 1.00 | ≈ 1.00 | ≈ 1.00 |
| PlainCNN1D | 0.999 | ≈ 1.00 | ≈ 1.00 | ≈ 1.00 |
| TCN | 1.000 | ≈ 1.00 | ≈ 1.00 | ≈ 1.00 |
| TinyTransformer | 0.991 | ≈ 0.99 | ≈ 1.00 | ≈ 0.98 |

### *5.2 Fault tolerance and the effect of sensor-dropout training*

Sensor-dropout training transforms tolerance to sensor loss (Table 3). With four of eight channels removed, overall AUC rises from 0.741 for the model trained without sensor-dropout (vanilla) to 0.982 with it (csd), at no measurable clean-data cost (both 1.000 at $k = 0$). Loss of a redundant arm or heart leaves performance essentially unchanged (≈ 1.000), confirming that the redundant components provide graceful degradation under internal failure. The two tiers of fault tolerance — sensor loss and component loss — are visualized in Fig. 3.

**Table 3.** Fault-tolerance ablation. Overall AUC for the model trained with sensor-dropout (csd) versus without it (vanilla), across loss conditions; Δ is csd minus vanilla.

| Condition | Vanilla | csd | Δ |
|---|---|---|---|
| Baseline (clean, k = 0) | 1.000 | 1.000 | +0.000 |
| 4 sensors lost | 0.741 | 0.982 | +0.241 |
| 7 sensors lost | 0.543 | 0.586 | +0.043 |
| 1 heart of 3 | 0.999 | 1.000 | +0.001 |
| 1 arm of 8 | 0.997 | 0.997 | +0.004 |

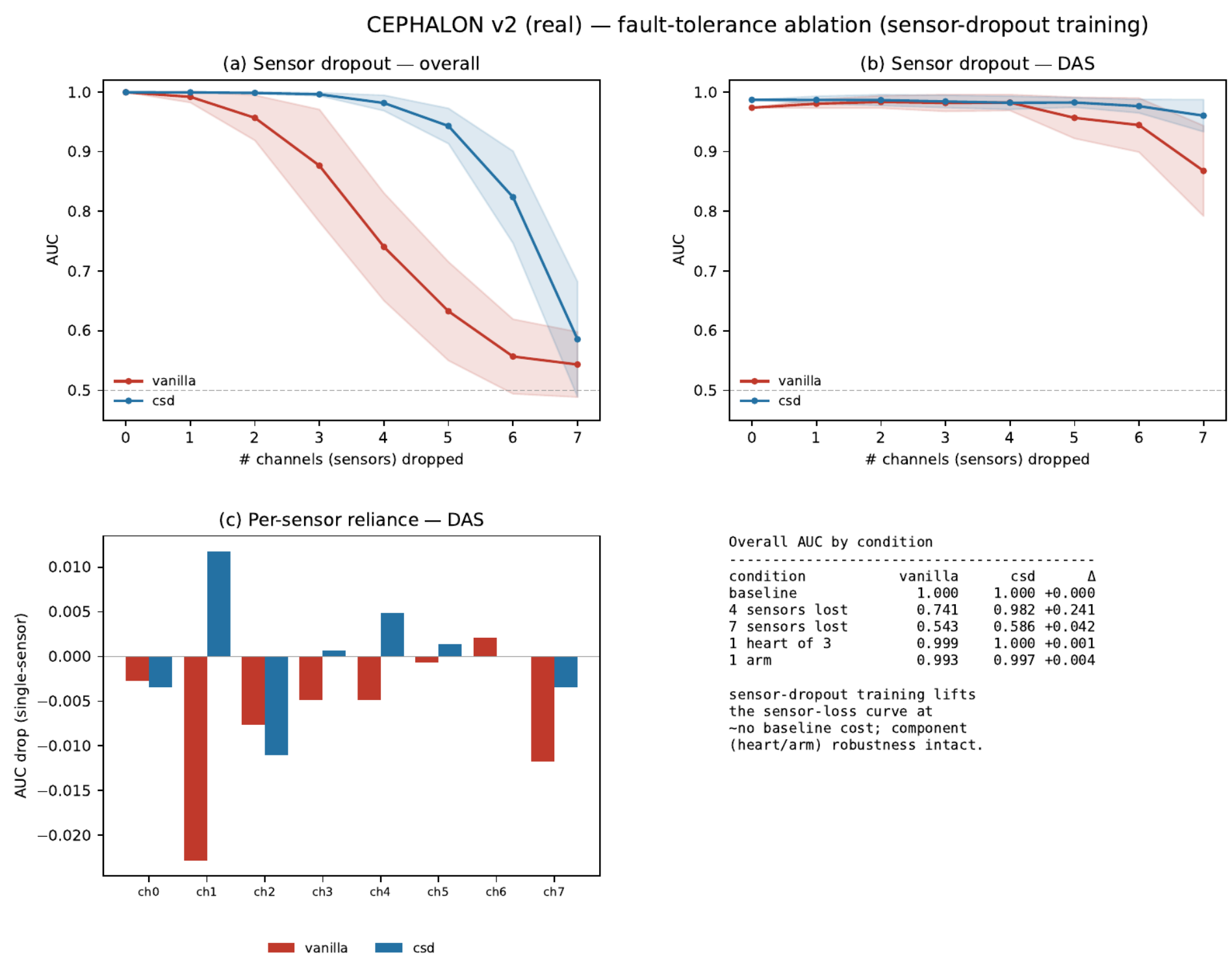

**Fig. 3.** Two-tier fault tolerance. Sensor-dropout training (csd) lifts a sensor-loss curve at no clean-data cost (4 sensors lost: 0.741 → 0.982; panels a, b), while per-channel reliance (panel c) shows the vanilla model leaning on individual channels that the csd model does not. Component (arm/heart) loss leaves AUC near 1.0.

### *5.3 Strong baselines: sensor-loss robustness is recipe-driven*

When the baselines are trained with the same sensor-dropout recipe, they are as robust as — or more robust than — CEPHALON under sensor loss (Table 4, Fig. 4a). With seven of eight channels removed, the TCN and PlainCNN1D retain overall AUC of 0.750 and 0.744 respectively, while CEPHALON falls to 0.586. Sensor-loss robustness is thus a property of the recipe, available to any model, not of the architecture. The second tier of fault tolerance — robustness to internal component loss — remains structurally unavailable to the baselines, which have no redundant pathways (Fig. 4b).

**Table 4.** Sensor-loss robustness. Overall test AUC for identically trained models at k = 0, 4, 7 dropped channels (random k-subsets).

| Model | k = 0 | k = 4 | k = 7 | Tier-2 |
|---|---|---|---|---|
| CEPHALON v2 (csd) | 1.000 | 0.982 | 0.586 | yes |
| PlainCNN1D | 0.999 | 0.986 | 0.744 | no |
| TCN | 1.000 | 0.993 | 0.750 | no |
| TinyTransformer | 0.991 | 0.899 | 0.698 | no |

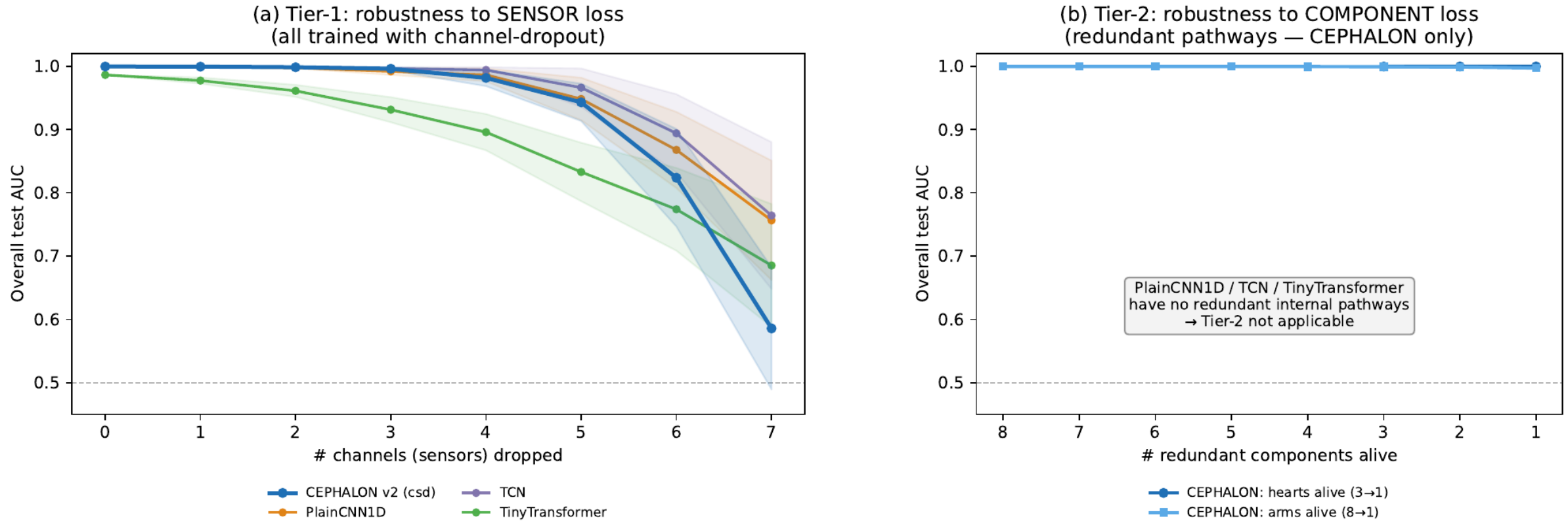

**Fig. 4.** Strong-baseline comparison. (a) Under sensor loss, identically trained simple models match or exceed CEPHALON. (b) Tier-2 robustness to internal component loss is structurally unavailable to baselines, which lack redundant pathways.

### *5.4 Low-SNR robustness (main result)*

The picture changes sharply under additive noise (Table 5, Fig. 5). As test SNR falls, the PlainCNN1D and TCN collapse abruptly — by −2.5 dB their overall AUC has fallen to 0.572 and 0.532 respectively — whereas CEPHALON degrades gracefully, retaining 0.939 at −2.5 dB and 0.899 at −5 dB; the TinyTransformer is intermediate (0.805 at −2.5 dB). The separation is consistent across all domains and is most pronounced for DAS, where CEPHALON retains 0.893 at −5 dB and 0.907 at −10 dB, while the convolutional baselines fall to chance (≈ 0.50) by −5 dB. This is the practically decisive axis for field monitoring, because the events that matter most for induced-seismicity catalogues are precisely the small, low-SNR ones.

**Table 5.** Low-SNR robustness. Overall AUC versus test SNR (mean over 3 noise seeds, white noise). DAS-panel values are given for the modality with the largest margin.

| SNR | CEPHALON | CNN | TCN | Transf. | DAS (CEPH) |
|---|---|---|---|---|---|
| clean / high | ≈ 1.00 | ≈ 1.00 | ≈ 1.00 | ≈ 0.99 | ≈ 1.00 |
| 0 dB | 0.970 | ≈ 0.93 | ≈ 0.82 | ≈ 0.89 | ≈ 0.98 |
| −2.5 dB | 0.939 | 0.572 | 0.532 | 0.805 | ≈ 0.95 |
| −5 dB | 0.899 | 0.512 | 0.502 | 0.698 | 0.893 |
| −10 dB | 0.738 | ≈ 0.50 | ≈ 0.50 | ≈ 0.53 | 0.907 |

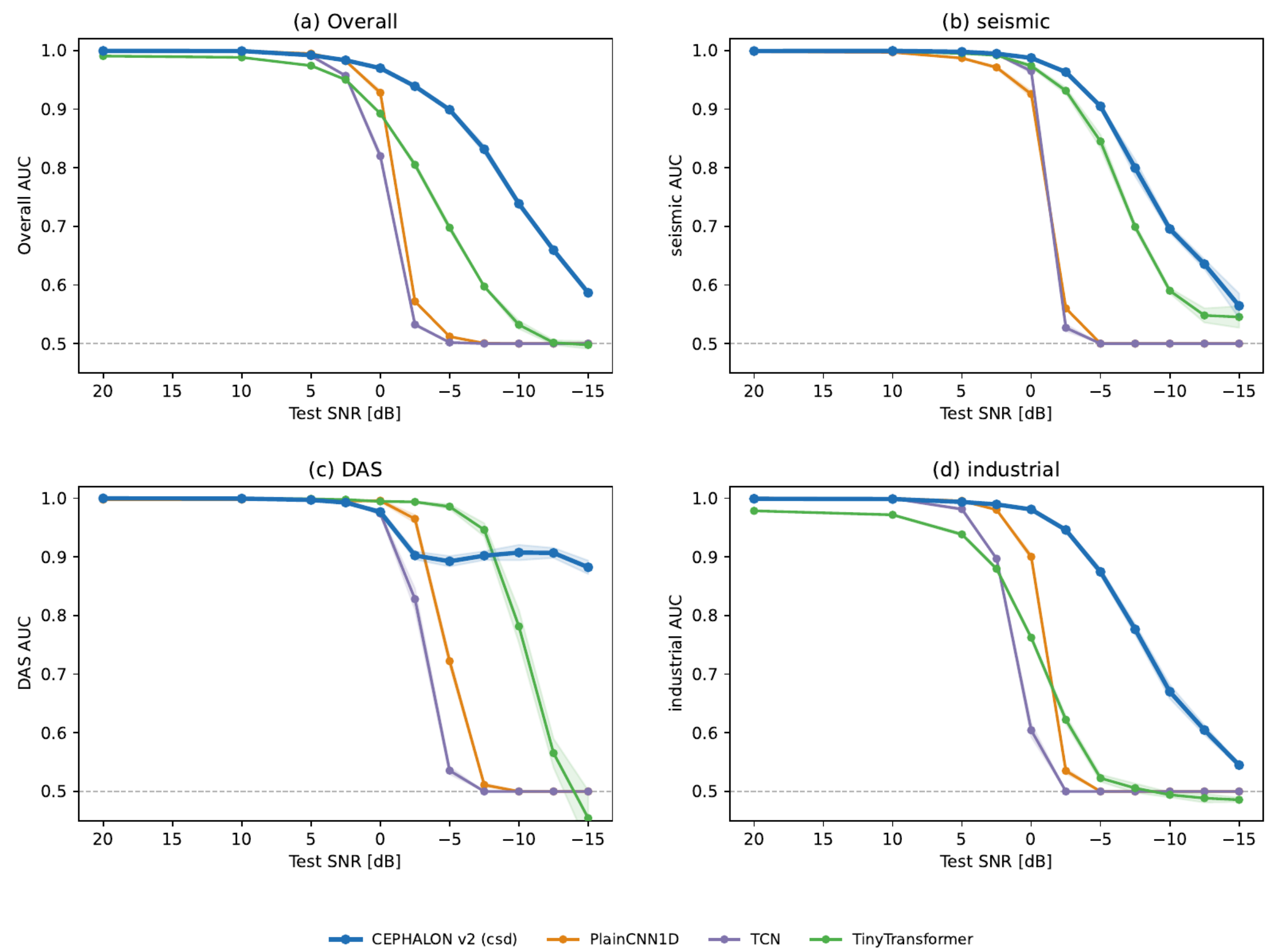

**Fig. 5.** Low-SNR robustness (mean ± s.d., 3 noise seeds; n = 150 DAS test windows), overall and per domain. CNN/TCN collapse near 0 dB; CEPHALON degrades gracefully. DAS panel (c) shows the largest margin (CEPHALON 0.91 at −10 dB versus baselines at chance).

### *5.5 Mechanism: recipe, not redundancy*

Two same-architecture ablations separate the contributions (Table 6, Fig. 6). First, disabling CEPHALON's internal redundancy at inference — running with only 1 of 8 arms and 1 of 3 hearts — reduces the low-SNR AUC only modestly and leaves it far above the baselines (0.853 versus 0.899 for the full model at −5 dB, against ≤ 0.51 for CNN/TCN). Because the model was trained to operate with arbitrary arm and heart subsets, this configuration is in-distribution; parallel redundancy therefore contributes only secondarily. Second, removing sensor-dropout from training (the otherwise identical vanilla model) collapses the low-SNR curve far more severely — from 0.970 to 0.808 at 0 dB, and from 0.899 to 0.603 at −5 dB — approaching the baselines. The dominant cause of the low-SNR robustness is thus the sensor-dropout training recipe (recipe-off: −0.30 AUC at −5 dB), with architectural redundancy a smaller, secondary contributor (redundancy-off: −0.05 AUC at −5 dB).

**Table 6.** Mechanism ablation. Overall AUC under low SNR for the full model versus redundancy-off and recipe-off variants; Δ is relative to the full model at −5 dB.

| Configuration | 0 dB | −5 dB | Δ (−5 dB) |
|---|---|---|---|
| Full CEPHALON (csd) | 0.970 | 0.899 | — |
| Redundancy-off (1 arm, 1 heart) | ≈ 0.96 | 0.853 | −0.05 |
| Recipe-off (vanilla, no sensor-dropout) | 0.808 | 0.603 | −0.30 |

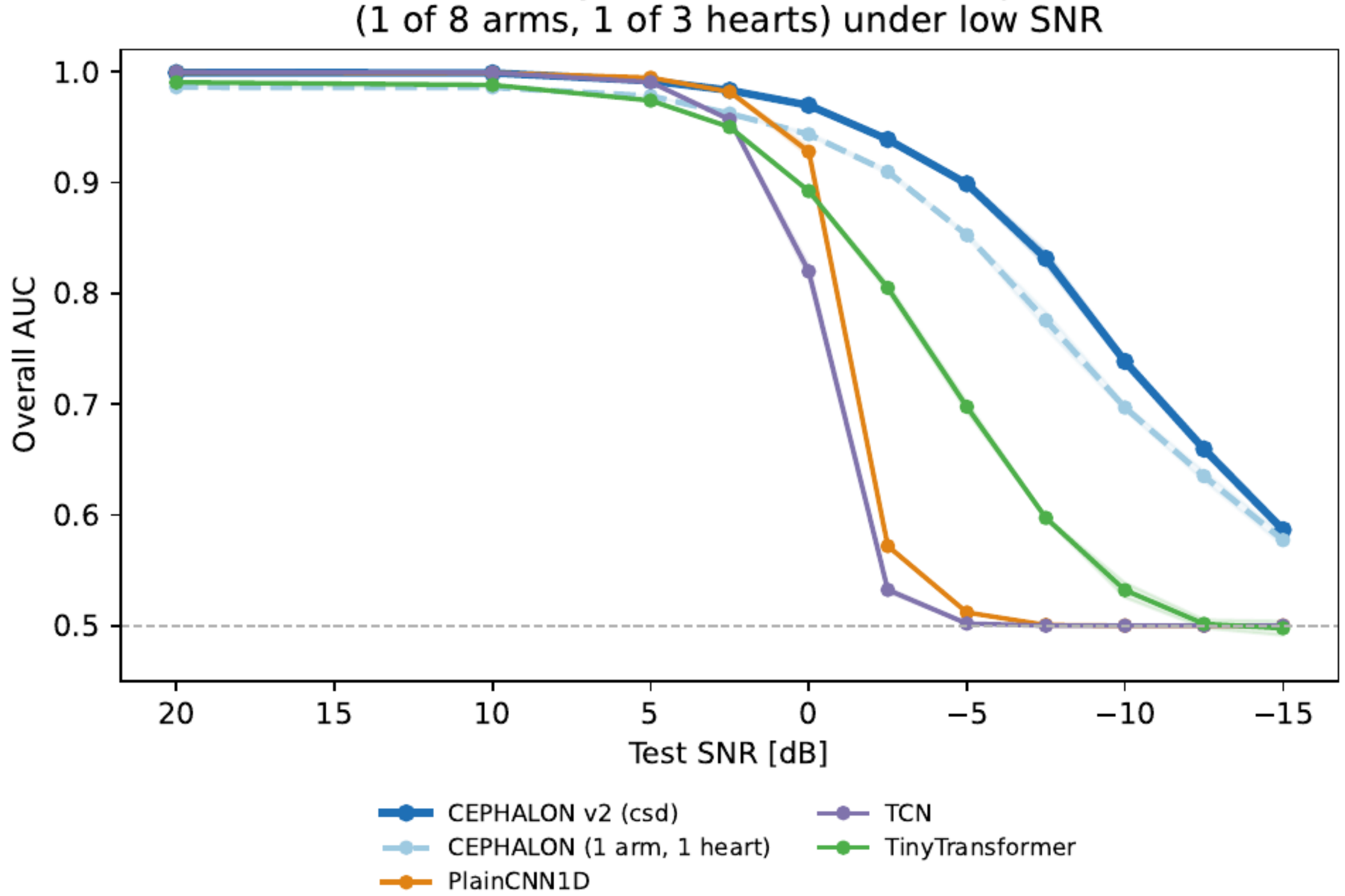

**Fig. 6.** Mechanism. Disabling internal redundancy at inference (1 arm, 1 heart; dashed) sits between a full model (solid) and baselines, so redundancy contributes only secondarily; removing a sensor-dropout recipe collapses a curve to baseline level, identifying a recipe as a dominant cause (see text and Table 6).

### *5.6 Cross-domain transfer (supporting)*

Zero-shot transfer across these physically distinct domains fails: leave-one-domain-out evaluation is at chance (overall AUC 0.241 for the held-out DAS domain). However, the multi-domain representation supports rapid few-shot adaptation (Table 7, Fig. 7). With only K = 5 labeled target examples per class, co-training with the source domains reaches DAS AUC 0.93, exceeding training on those examples alone (0.89), and the two converge to the in-domain ceiling (≈ 0.974) by about K = 50. Generalization here is few-shot, not zero-shot, and we report the null zero-shot result rather than omit it.

**Table 7.** Cross-domain few-shot transfer to DAS. DAS test AUC for transfer (source + K target) versus scratch (K target only); zero-shot is leave-one-domain-out.

| K (per class) | Transfer (source + K) | Scratch (K only) |
|---|---|---|
| 0 (zero-shot / LODO) | 0.241 | — |
| 5 | 0.93 | 0.89 |
| ≥ 50 | → ceiling (≈ 0.974) | → ceiling (≈ 0.974) |

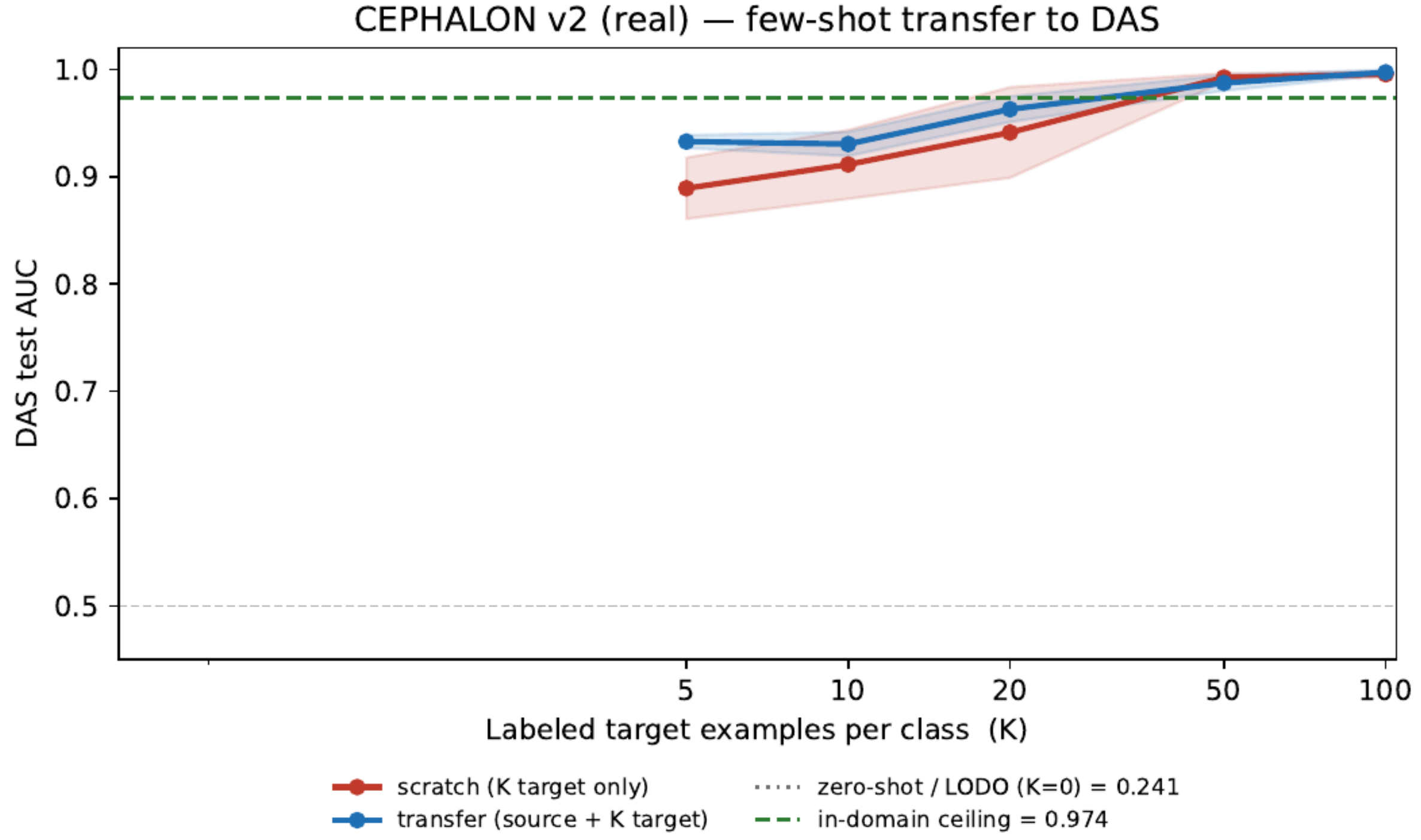

**Fig. 7.** Cross-domain transfer to DAS. Zero-shot (K = 0) is at chance; few-shot co-training adapts rapidly and exceeds training on K target examples alone, converging to an in-domain ceiling.

## 6. Discussion

### *6.1 Two robustness axes with distinct causes*

The two robustness axes are distinct and have distinct causes. Robustness to sensor loss is conferred by the training recipe and is available to any of the detectors we tested; under clean sensor loss a lightweight convolutional network trained with sensor-dropout matches or exceeds the redundant architecture. Robustness to low SNR is also recipe-driven, but it is not shared by the simple convolutional baselines despite their using the same recipe, which points to an interaction between the recipe and the more expressive trunk. Crucially, the architecture's most conspicuous feature — its parallel redundancy — is not the source of the low-SNR advantage, as the inference-time redundancy ablation shows. Conflating the two robustness types, or crediting architecture for what the recipe delivers, would misattribute the effect and would mislead practitioners about where to invest.

### *6.2 Why channel sensor-dropout transfers to additive noise*

Why should sensor-dropout, a channel-masking augmentation, help against additive noise, which is a different corruption? Sensor-dropout is a structured input perturbation that discourages reliance on any single channel and encourages distributed, redundant feature use. A detector trained to classify correctly when arbitrary channels are zeroed cannot place excessive weight on any one channel, and a representation that is robust to the removal of channels is also less easily dominated by noise injected into any subset of them. The effect therefore appears to transfer to a corruption it never saw during training — a form of free, recipe-induced regularization that aligns with the broader finding that targeted augmentation improves corruption robustness more reliably than architectural change (Hendrycks and Dietterich, 2019).

### *6.3 Recipe versus architecture, and parameter efficiency*

The practical corollary of the mechanism result is favourable to lightweight deployment. The redundant CEPHALON architecture is substantially more expensive than the convolutional baselines in parameters and inference cost, yet a large fraction of the robustness that matters in the field — all of the sensor-loss robustness, and a meaningful share of the low-SNR robustness — can be obtained from a simple, inexpensive detector simply by training it with sensor-dropout. The redundant architecture still earns two distinct benefits that the baselines cannot: graceful degradation under internal component loss (Tier-2 fault tolerance), and the residual low-SNR margin that is largest precisely on DAS. Whether that residual margin justifies the additional cost is an engineering decision that depends on the deployment; the contribution of this paper is to make the trade-off explicit and measurable rather than to assume that the heavier model is always preferable.

### *6.4 Robustness under colored (pink, 1/f) noise*

Because real field noise is colored and non-stationary rather than white, we repeated the low-SNR experiment with pink (1/f) noise (Appendix B). The CEPHALON advantage at low SNR persists under pink noise, most clearly below −5 dB, and the mechanism ablation again places the redundancy-off curve near the full model. The low-SNR robustness is therefore not an artifact of the white-noise model, although fully realistic evaluation against recorded DAS noise remains important future work.

### *6.5 Implications for field monitoring*

For field monitoring of CCS and geothermal induced seismicity the implication is direct. Acquisition systems lose channels and operate at low SNR as a matter of course, and the small events that dominate induced-seismicity catalogues are exactly those most vulnerable to noise. Our results indicate that a substantial part of the robustness that matters in such networks can be obtained from a simple detector trained with sensor-dropout, and that the training recipe deserves at least as much attention as the architecture. The residual advantage of the more expressive model is real, is largest on DAS — the modality most relevant to dense borehole monitoring — and can be reserved for deployments where the additional cost is acceptable.

### *6.6 Limitations and future directions*

Several limitations bound these conclusions. The detection task is intrinsically easy on clean data; all of our conclusions live in the low-SNR and sensor-loss tails, where the contrasts are large. The DAS test set, although doubled to 497 events (150 test windows), remains the smallest domain, and its low-SNR curve is correspondingly the most variable. Additive Gaussian noise is a convenient but imperfect proxy for field noise; the pink-noise check mitigates but does not eliminate this concern, and evaluation against recorded DAS noise is the most important next step.

The most extreme sensor-loss condition (one of eight channels alive) is near-degenerate for every model and should be read with caution. Finally, the few-shot transfer result is established for transfer to DAS; extending it to the other held-out domains, and characterizing the fine-tuning regime, would strengthen the multi-domain claim.

## 7. Conclusion

On a unified benchmark of three physically distinct real datasets — Hi-net seismic, Utah FORGE 2024 borehole DAS, and MAFAULDA industrial vibration — we separated two robustness properties that are usually conflated and identified their causes. Sensor-loss robustness is a property of the sensor-dropout training recipe and is shared by simple baselines. Low-SNR robustness, the practically decisive property for field monitoring, is also recipe-driven rather than a product of the architecture's parallel redundancy; this is established by same-architecture ablations in which removing the recipe collapses the advantage while removing the redundancy does not. The result reframes a common assumption in monitoring practice: robustness that is routinely credited to architecture is, here, bought far more cheaply by the training recipe. We release the complete, numbered, reproducible pipeline so that every figure and table can be regenerated.

## Appendix A. DAS ingestion quality control

Quality-control diagnostics for the Utah FORGE 2024 DAS ingestion are shown in Fig. A1: event/noise window-amplitude separation, the magnitude distribution, the magnitude–depth relationship, and per-channel quality over the illuminated band (channels 1116–1493). The window-amplitude separation confirms that the amplitude cue preserved by the noise-relative normalization is the dominant discriminator between events and paired noise windows.

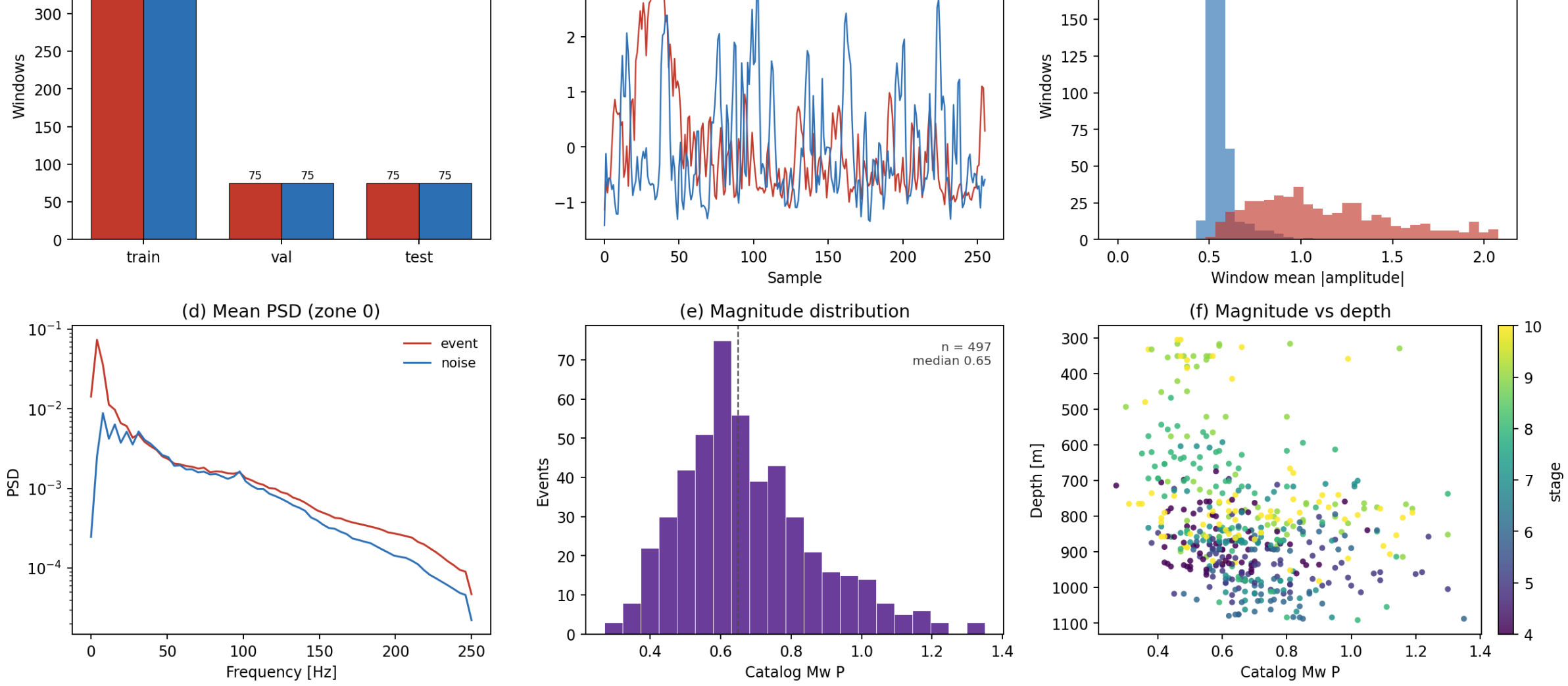

**Fig. A1.** FORGE 2024 DAS ingestion quality control: event/noise window-amplitude separation, magnitude distribution, magnitude–depth, and per-channel quality over an illuminated band (channels 1116–1493).

## Appendix B. Low-SNR robustness under pink (1/f) noise

To check that the low-SNR advantage is not specific to white noise, the main low-SNR experiment was repeated with pink (1/f) noise. The CEPHALON advantage persists (Fig. B1), most clearly below −5 dB, and the redundancy-off mechanism ablation again sits near the full model (Fig. B2), consistent with the recipe being the dominant cause under colored as well as white noise.

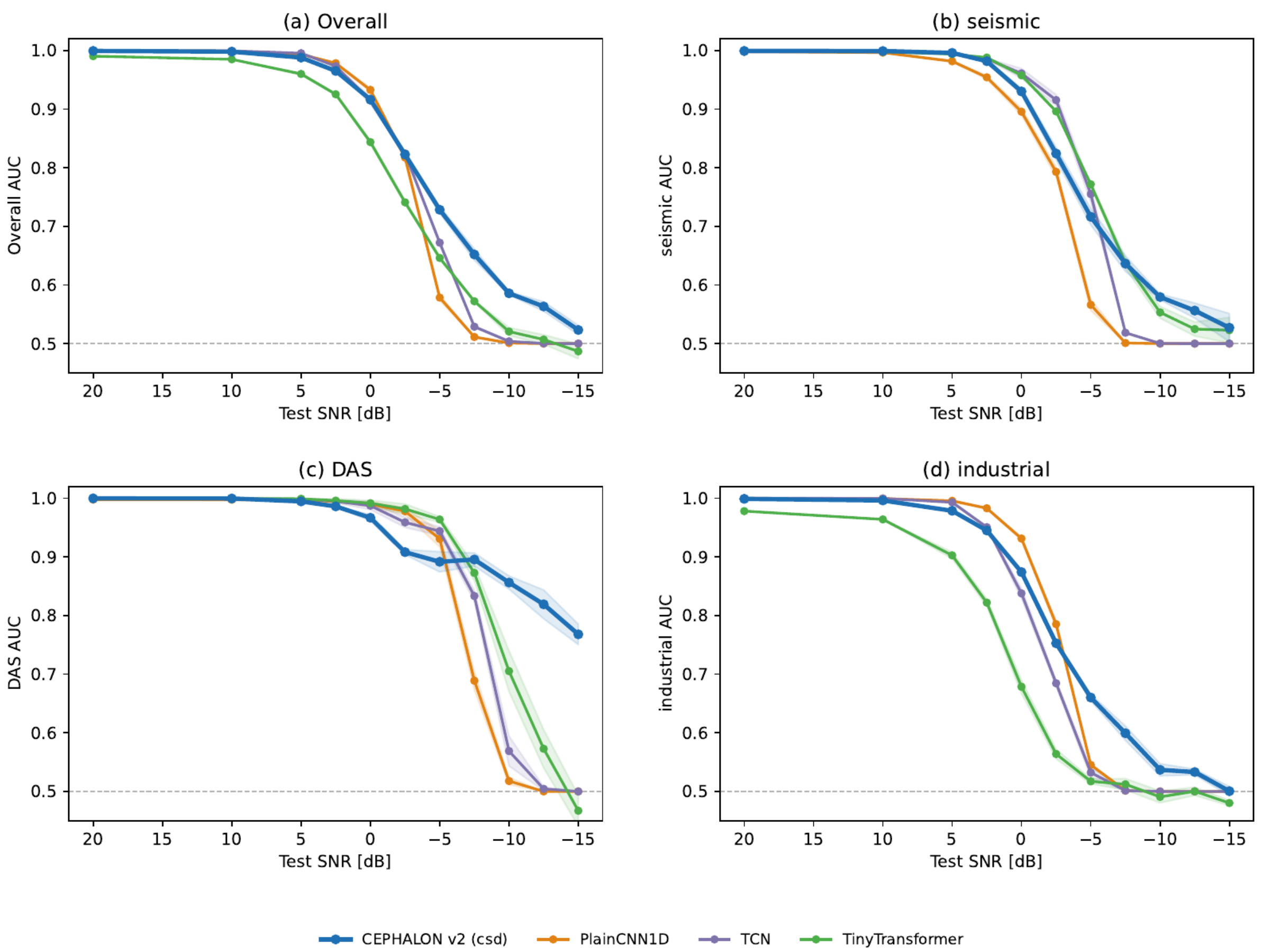

**Fig. B1.** Low-SNR robustness under pink (1/f) noise (mean ± s.d., 3 seeds), overall and per domain. A CEPHALON advantage at low SNR persists with colored noise, most clearly below −5 dB.

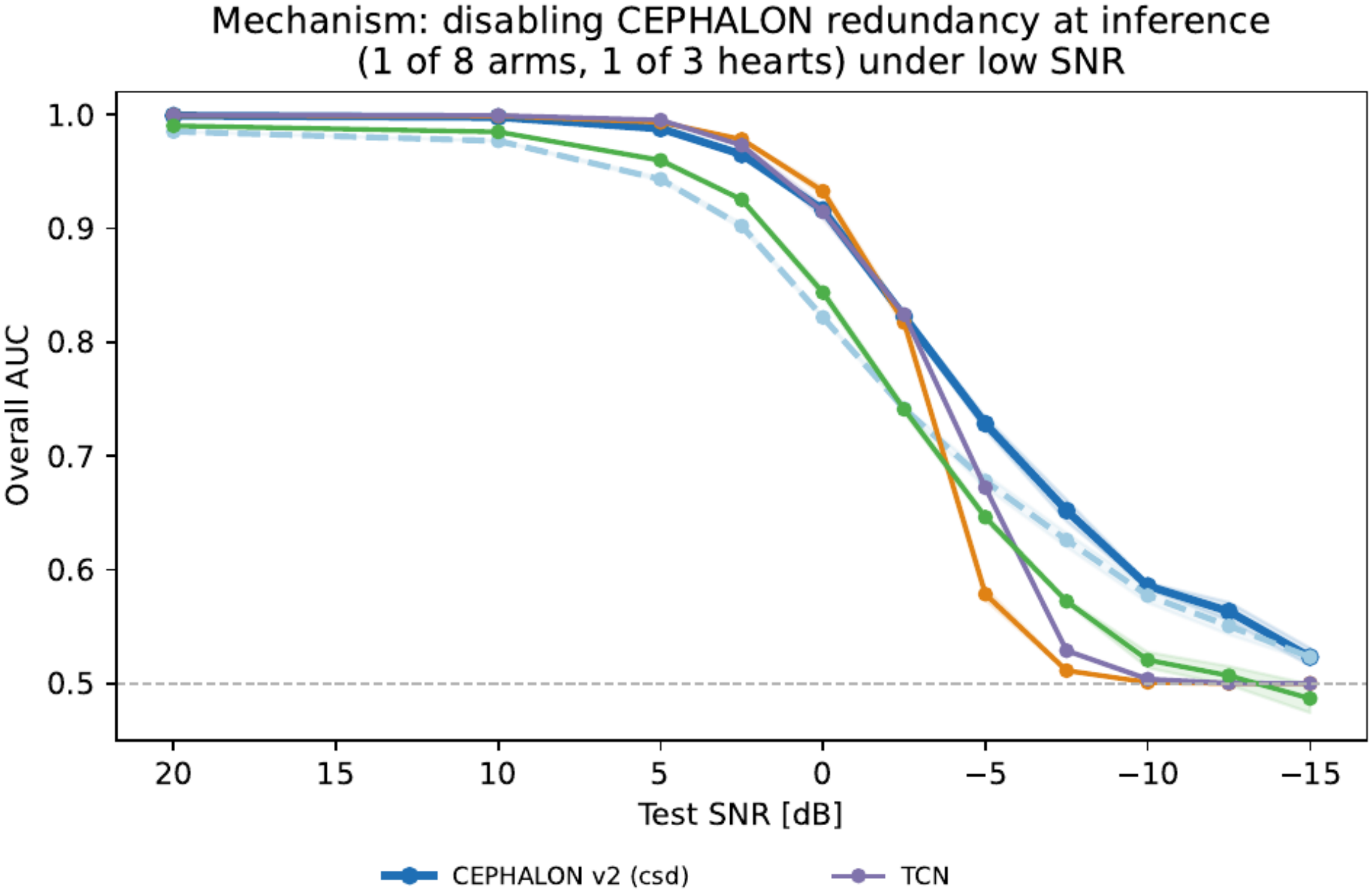

**Fig. B2.** Mechanism under pink noise: redundancy-off (1 arm, 1 heart) again sits near a full model, consistent with a recipe being a dominant cause.

## Code and data availability

All experiments are produced by a numbered pipeline (ingestion, multi-domain assembly, training, fault-tolerance and low-SNR evaluation, baseline training, and figure generation) run in a uv environment. The code, a documented README with installation and usage instructions, a license, dependency and computational-requirement documentation, and a synthetic test case sufficient to reproduce the main results are provided in a public repository (https://github.com/ISAO9/cephalon-v2), archived at Zenodo (https://doi.org/10.5281/zenodo.20995392). All datasets are public: Hi-net (NIED), Utah FORGE 2024 DAS in monitoring well 16B (Geothermal Data Repository, doi:10.15121/2479771), and the MAFAULDA machinery-fault database.

## Declaration of competing interest

The author declares that he has no known competing financial interests or personal relationships that could have appeared to influence the work reported in this paper.

## Declaration of generative AI and AI-assisted technologies in the manuscript preparation process

During the preparation of this work the author used a large language model to assist with manuscript drafting and language editing. After using this tool, the author reviewed and edited the content and takes full responsibility for the content of the published article.

## CRediT authorship contribution statement

Isao Kurosawa: Conceptualization, Methodology, Software, Validation, Formal analysis, Investigation, Data curation, Writing – original draft, Writing – review and editing, Visualization.

## Acknowledgments

The author thanks the NIED Hi-net, the Utah FORGE project and Neubrex, and the SMT/COPPE MAFAULDA team for making their data publicly available.

## Funding

This research did not receive any specific grant from funding agencies in the public, commercial, or not-for-profit sectors.

## References

Ajo-Franklin, J., et al., 2024. Utah FORGE: distributed acoustic sensing (DAS) microseismic event triggers (waveforms), 16A–16B stimulations, April 2024. Geothermal Data Repository (Utah FORGE Project 3-2417). [confirm submission DOI on gdr.openei.org]

Bai, S., Kolter, J.Z., Koltun, V., 2018. An empirical evaluation of generic convolutional and recurrent networks for sequence modeling. arXiv:1803.01271.

Bradley, A.P., 1997. The use of the area under the ROC curve in the evaluation of machine learning algorithms. Pattern Recognition 30, 1145–1159. https://doi.org/10.1016/S0031-3203(96)00142-2.

DeVries, T., Taylor, G.W., 2017. Improved regularization of convolutional neural networks with cutout. arXiv:1708.04552.

Ellsworth, W.L., 2013. Injection-induced earthquakes. Science 341, 1225942. https://doi.org/10.1126/science.1225942.

Ganin, Y., Ustinova, E., Ajakan, H., Germain, P., Larochelle, H., Laviolette, F., Marchand, M., Lempitsky, V., 2016. Domain-adversarial training of neural networks. Journal of Machine Learning Research 17, 1–35.

Grigoli, F., Cesca, S., Priolo, E., Rinaldi, A.P., Clinton, J.F., Stabile, T.A., Dost, B., Garcia Fernandez, M., Wiemer, S., Dahm, T., 2017. Current challenges in monitoring, discrimination, and management of induced seismicity related to underground industrial activities: a European perspective. Reviews of Geophysics 55, 310–340. https://doi.org/10.1002/2016RG000542.

Hendrycks, D., Dietterich, T., 2019. Benchmarking neural network robustness to common corruptions and perturbations. International Conference on Learning Representations.

Kingma, D.P., Ba, J., 2015. Adam: a method for stochastic optimization. International Conference on Learning Representations.

Kipf, T.N., Welling, M., 2017. Semi-supervised classification with graph convolutional networks. International Conference on Learning Representations.

Lindsey, N.J., Martin, E.R., 2021. Fiber-optic seismology. Annual Review of Earth and Planetary

Sciences 49, 309–336. https://doi.org/10.1146/annurev-earth-072420-065213.

Marins, M.A., Ribeiro, F.M.L., Netto, S.L., da Silva, E.A.B., 2018. Improved similarity-based modeling for the classification of rotating-machine failures. Journal of the Franklin Institute 355, 1913–1930. https://doi.org/10.1016/j.jfranklin.2017.07.038.

Mousavi, S.M., Beroza, G.C., 2022. Deep-learning seismology. Science 377, eabm4470. https://doi.org/10.1126/science.abm4470.

Mousavi, S.M., Ellsworth, W.L., Zhu, W., Chuang, L.Y., Beroza, G.C., 2020. Earthquake transformer—an attentive deep-learning model for simultaneous earthquake detection and phase picking. Nature Communications 11, 3952. https://doi.org/10.1038/s41467-020-17591-w.

Okada, Y., Kasahara, K., Hori, S., Obara, K., Sekiguchi, S., Fujiwara, H., Yamamoto, A., 2004. Recent progress of seismic observation networks in Japan—Hi-net, F-net, K-NET and KiK-net. Earth, Planets and Space 56, xv–xxviii. https://doi.org/10.1186/BF03353076.

Perol, T., Gharbi, M., Denolle, M., 2018. Convolutional neural network for earthquake detection and location. Science Advances 4, e1700578. https://doi.org/10.1126/sciadv.1700578.

Ribeiro, F.M.L., 2018. MAFAULDA: machinery fault database. SMT/COPPE, Federal University of Rio de Janeiro. http://www02.smt.ufrj.br/~offshore/mfs/page_01.html.

Ross, Z.E., Meier, M.-A., Hauksson, E., Heaton, T.H., 2018. Generalized seismic phase detection with deep learning. Bulletin of the Seismological Society of America 108, 2894–2901. https://doi.org/10.1785/0120180080.

Shazeer, N., Mirhoseini, A., Maziarz, K., Davis, A., Le, Q., Hinton, G., Dean, J., 2017. Outrageously large neural networks: the sparsely-gated mixture-of-experts layer. International Conference on Learning Representations.

Srivastava, N., Hinton, G., Krizhevsky, A., Sutskever, I., Salakhutdinov, R., 2014. Dropout: a simple way to prevent neural networks from overfitting. Journal of Machine Learning Research 15, 1929–1958.

Utah FORGE GDR, 2024. Neubrex well 16B(78)-32 DAS data—April 2024. Geothermal Data Repository. https://doi.org/10.15121/2479771.

van den Ende, M., Lior, I., Ampuero, J.-P., Sladen, A., Ferrari, A., Richard, C., 2021. A self-supervised deep learning approach for blind denoising and waveform coherence enhancement in distributed acoustic sensing data. IEEE Transactions on Neural Networks and Learning Systems. https://doi.org/10.1109/TNNLS.2021.3132832.

Vaswani, A., Shazeer, N., Parmar, N., Uszkoreit, J., Jones, L., Gomez, A.N., Kaiser, L., Polosukhin, I., 2017. Attention is all you need. Advances in Neural Information Processing Systems.

Zhu, W., Beroza, G.C., 2019. PhaseNet: a deep-neural-network-based seismic arrival-time picking method. Geophysical Journal International 216, 261–273. https://doi.org/10.1093/gji/ggy423.

Zhu, W., Biondi, E., Li, J., Yin, J., Ross, Z.E., Zhan, Z., 2023. Seismic arrival-time picking on distributed acoustic sensing data using semi-supervised learning. Nature Communications 14, 8192. https://doi.org/10.1038/s41467-023-43355-3.